\documentclass[submitted, 12pt]{elsarticle}
\usepackage{bbold}
\usepackage{graphicx}
\usepackage{amssymb}
\usepackage{amsmath}
\usepackage{empheq}
\usepackage{amsthm}
\usepackage{siunitx}
\usepackage{float}

\newcommand{\bu}{\boldsymbol{u}}
\newcommand{\bPsi}{\boldsymbol{\Psi}}

\usepackage{subfig}

\usepackage{amssymb}
\usepackage{amsthm}
\usepackage{amsmath}
\usepackage{color}
\usepackage{lineno}
\usepackage{hyperref}
\hypersetup{
    colorlinks=true,
    linkcolor=blue,
    filecolor=magenta,      
    urlcolor=cyan,
    pdftitle={Overleaf Example},
    pdfpagemode=FullScreen,
    }

\DeclareSymbolFontAlphabet{\mathrm}{operators}

\graphicspath{ {Figures/} }

\let\today\relax
\makeatletter
\def\ps@pprintTitle{%
    \let\@oddhead\@empty
    \let\@evenhead\@empty
    \def\@oddfoot{\footnotesize\itshape
         {} \hfill\today}%
    \let\@evenfoot\@oddfoot
    }
\makeatother

\begin{document}

\begin{frontmatter}

\title{The Elastic Spiral Phase Pipe \footnote{This is the submitted version of \textit{G.J. Chaplain, and J.M. De Ponti, The Elastic Spiral Phase Pipe, Journal of Sound and Vibration, 523, 116718 (2022)}. The final publication is available at \href{https://doi.org/10.1016/j.jsv.2021.116718}{https://doi.org/10.1016/j.jsv.2021.116718}}}


\author[label1,label2]{G.~J. Chaplain$^{*}$}
\ead{g.j.chaplain@exeter.ac.uk}
\address[label1]{Department of Mathematics, Imperial College London, 180 Queen's Gate, South Kensington, London SW7 2AZ}
\address[label2]{Electromagnetic and Acoustic Materials Group, Department of Physics and Astronomy, University of Exeter, Exeter EX4 4QL, United Kingdom}
\author[label3]{J.M. De Ponti}%
\address[label3]{Department of Civil and Environmental Engineering, Politecnico di Milano, Piazza Leonardo da Vinci, 32, 20133 Milano, Italy }%

\begin{abstract}
We design a device for the passive mode conversion of guided, axisymmetric, ultrasonic waves in hollow elastic pipes into arbitrary non-axisymmetric flexural waves that have a constant angular profile along the pipe axis. To achieve this we create an elastic analogue to optical spiral phase plates - the elastic spiral phase pipe.  Three possible configurations of the elastic spiral phase pipe are presented which allow the efficient generation of non-axisymmetric flexural waves from an axisymmetric, longitudinal forcing. The theory leverages the dispersive nature of the guided elastic waves that are supported in pipes through a defined relative refractive index. As such we include a spectral collocation method used to aid the design of the elastic spiral phase pipe that is corroborated with numerical simulations and then experimentally verified.
\end{abstract}


\end{frontmatter}

\section{\label{sec:intro} Introduction}
The generation, manipulation and inspection of (ultrasonic) guided waves along structures has played an important role in the development of non-destructive testing (NDT) techniques. These techniques enable elastic material properties to be determined and have uses in the inspection of welding defects, cracks and adhesive joints, all crucially without causing damage to the sample. Given the ubiquity of large pipe networks across many sectors of industry, from petrochemical to power generation, these techniques are particularly pertinent for assessing the structural integrity of pipe structures \cite{krautkramer2013ultrasonic}. To utilise NDT techniques effectively it is therefore essential to have efficient means of exciting the guided ultrasonic waves, which requires analysis of the types of waves supported by a given structure.

There is a long and rich history of the analysis of propagating guided waves in hollow cylinders. The first general solution for harmonic waves
propagating in an infinitely long hollow cylinder was derived by Gazis \cite{gazis1959a,gazis1959b}, who showed there are three families of modes: longitudinal, torsional and flexural. Following the conventional labelling system of Silk and Bainton \cite{silk1979propagation}, these are classed as $L(m,n)$, $T(m,n)$ and $F(m,n)$ respectively, where $m$ denotes the circumferential order and $n$ is the group order, or labelling index, with each mode possessing unique dispersive properties. The mode shapes for which $m = 0$ are axisymmetric i.e. their angular profile is constant. Therefore only the $T(0,n)$ are of pure torsion, and the $L(0,n)$ are essentially axial extensional modes, with varying radial profiles \cite{LOWE20011551}. The flexural modes $F(m > 0,n)$ are non-axisymmetric modes whose mode shapes vary sinusoidally in the circumferential direction. Alternative naming conventions only use $L(m,n)$ and $T(m,n)$ where it is understood that only $m = 0$ corresponds to axisymmetric modes with $T(m > 0,n)$ being flexural modes \cite{nishino2011experimental,rose2014ultrasonic}.  

The axisymmetric guided wave modes $L(0,1)$, $L(0,2)$ and $T(0,1)$ can be easily excited over typical pipe sizes and frequencies \cite{lowe2015inspection,rose2002standing}. Most commonly used in pipe inspection are the $L(0,2)$ and $T(0,1)$ modes, due to their ability to be efficiently excited and their dispersive properties \cite{alleyne1996excitation,alleyne2001rapid,lowe1998defect,niu2019excitation}; the $L(0,2)$ mode has the highest phase velocity and is only weakly dispersive, whilst the $T(0,1)$ is non-dispersive across the whole frequency spectrum. These modes are sensitive to circumferential defects and had have much success in the field of non-destructive testing and evaluation \cite{Alleyne98,Lowe98,demma2003reflection,nishino2015investigation}. Higher order torsional modes, e.g. $T(0,2)$, have also have uses in NDT but require more complex electromagnetic acoustic transducers (EMATs) for excitation \cite{nakamura2017emat}.

However, the $L(0,2)$ modes are insensitive to axial cracks,
and, despite the particle displacement for the $T(0,1)$ mode being perpendicular to the plane of propagation, this mode requires deep cracks for detectable signals \cite{kwun2008detection,ratassepp2010scattering}. There has therefore been several efforts to investigate the potential of guided non-axisymmetric flexural modes for pipe inspection \cite{shin1998guided,tang2017excitation}. These modes too have their own complications due to their dispersive nature and mode shapes that change along the direction of propagation; the circumferential distribution of particle displacements (the angular profile) for non-axisymmetric guided waves is complex, and changes with propagation distance, frequency, and mode \cite{ditri1992excitation}. This is due to the unique dispersive properties of the groups of modes present; the differences in phase velocities result in a superposition leading to a varying angular profile \cite{li2001excitation}. For such dispersive wave packets there is a limit on the long range resolution in NDT applications \cite{wilcox2001effect}. The non-trivial mode shapes of the flexural modes require complex arrangements of transducers (e.g comb arrays and non-axisymmetric partial loading \cite{shin1999guided,li2001excitation,Sun2005,li2006natural,tang2017excitation}) for their excitation, or by exploiting mode conversion from non-axisymmetric defects \cite{demma2003reflection} or bends \cite{demma2001mode,nishino2011experimental}. In order to tune the angular profile Li and Rose \cite{li2002angular} developed a circumferential phased array for implementing a circumferential scan with focused guided wave beams, requiring careful calibration of the transducer elements.
\begin{figure}[h!]
    \centering
    \includegraphics[width = \textwidth]{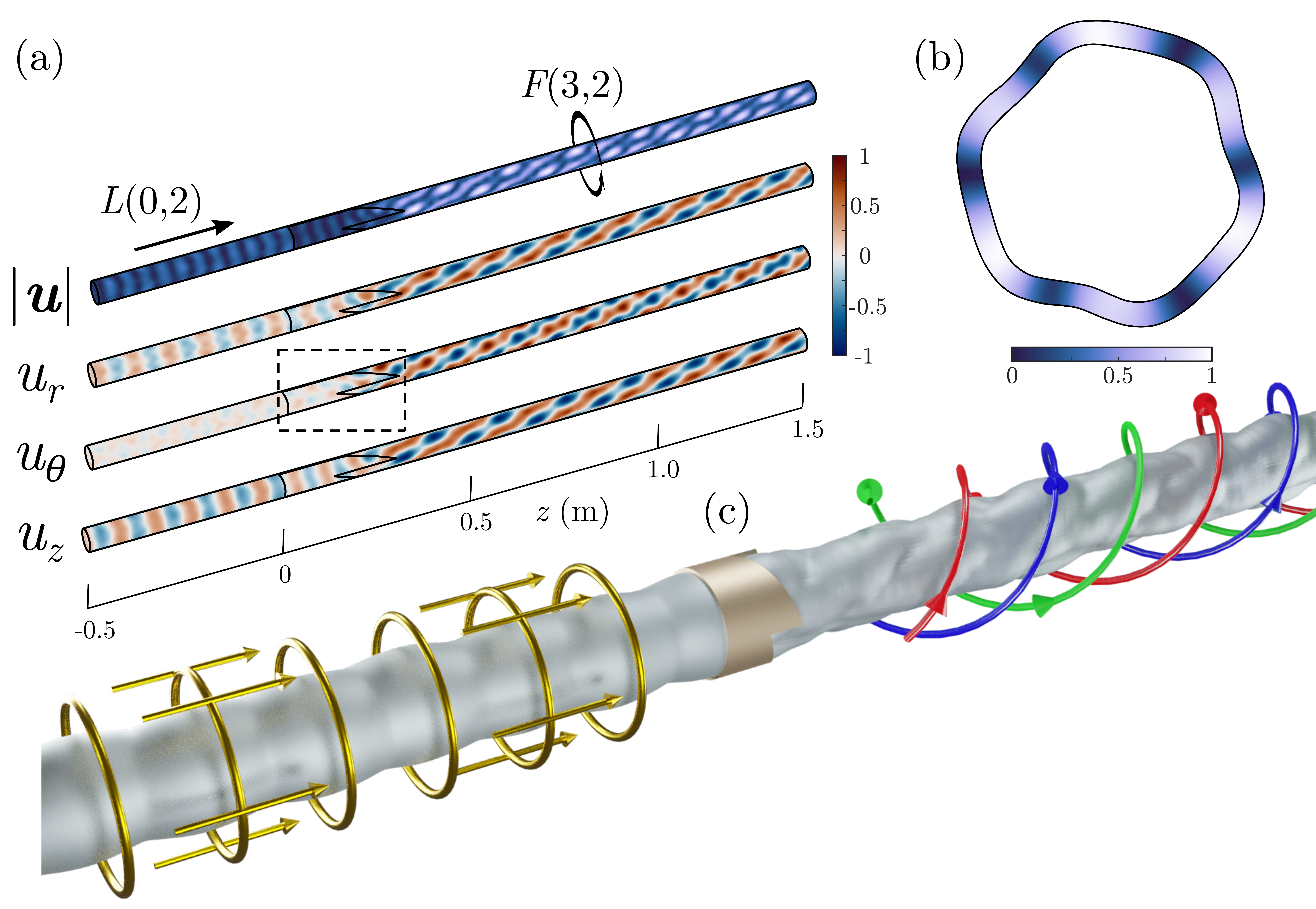}
    \caption{\textbf{The Elastic Spiral Phase Pipe:} (a) FEM frequency domain simulations showing normalised total solid displacement, $|\bu|$, and the normalised radial, angular and axial displacements $u_r$, $u_{\theta}$, $u_z$. A harmonic axisymmetric source is used. The position of the eSPP is shown in the dashed rectangle. Absorbing boundary conditions are used on the right end of the pipe. (b) Deformed cross sectional solid displacement field after the eSPP conversion (at position of circular arrow in (a)). (c) Artists impression of passive conversion of $L(0,2)$ to $F(3,2)$ in the steel pipe. Throughout this article perceptually uniform colour maps are used \cite{crameri2020misuse}.}
    \label{fig:hero}
\end{figure} 

We are motivated by the prospect of addressing the difficulties of exciting non-axisymmetric flexural modes in pipe structures and seek to generate them by passive mode conversion. Conversion of compressional to torsional motion has been studied in both dynamic \cite{tsujino1992ultrasonic} and static \cite{frenzel2017three} regimes. Here we focus specifically on the conversion of compressional waves to flexural waves. The proposed devices then have the advantages of (i) removing the need for complex transducer arrangements, replacing these with a \textit{single, axial source} (which can be point-like); (ii) exciting (theoretically) a single, desired flexural mode, rather than a group of modes; and (iii) ensuring only one `handedness' (or chirality) of the desired flexural modes is present (conventional excitation methods simultaneously excite both clockwise and anti-clockwise modes). We do so by taking inspiration from phased array effects, notably from the field of optics, and here employ a spatial inhomogeneity that can be incorporated into the pipe structure in one of three configurations. To demonstrate this capability, in Figure~\ref{fig:hero} we introduce the elastic spiral phase pipe (eSPP) - an elastic analogue to the optical spiral phase plate (oSPP) \cite{beijersbergen1994helical}.

Figure~\ref{fig:hero}(a) shows motivational results of a frequency domain finite element method (FEM) simulation displaying capabilities of mode converting a purely $L(0,2)$ mode almost completely into the (arbitrarily chosen) $F(3,2)$ mode. We show the normalised total solid displacement, $|\bu|$, and the normalised radial, angular and axial displacements $u_r$, $u_{\theta}$, $u_z$ respectively in a steel pipe of length $2.0~\si{\meter}$, $5~\si{\centi\meter}$ inner diameter and thickness $4.5~\si{\milli\meter}$. Here the $L(0,2)$ mode is excited with an axial forcing at $59.7~\si{\kilo\hertz}$ from the left. At the centre of the pipe is an eSPP (position marked by dashed rectangle in $u_\theta$ plot and highlighted by the bronzed region in the artists impression in Figure~\ref{fig:hero}(c)). This component is a hollowed circular helicoid (same diameter and thickness of the steel pipe) with a specifically designed azimuthally varying length profile that ensures efficient coupling to the $F(3,2)$ mode in the steel pipe at this design frequency. The azimuthal profile depends on the ratio of the speeds of the incident wave in the eSPP and the speed of desired mode to be excited after the device (in this illustrative example a tungsten join configuration is used - see Section~\ref{sec:Sec2}). The passive phase change endowed to the incident $L(0,2)$ as it traverses the eSPP results in mode conversion to a flexural $F(3,2)$ mode on the other side of the eSPP device. This is highlighted by the rotational displacement field $u_{\theta}$ in Figure~\ref{fig:hero}(a); before the device there is no circumferential displacement, whilst afterwords there is significant rotational motion. Figure~\ref{fig:hero}(b) shows a cross section after the device (at the position of the circular arrow in (a)), which shows the characteristic field distribution of the $F(3,2)$ mode. Figure~\ref{fig:hero}(c) shows an artists impression of the mode conversion effect: to the left of the eSPP an incident $L(0,2)$ mode is present (which is easily excited by an axial forcing), represented by the circular phase fronts shown in gold. After the device, helical phase fronts are shown consistent with the excitation of the \textit{single}, left-handed flexural mode $F(3,2)$. We remark that it is only the phase front of the guided wave which is helical in nature, not the hollow cylinder itself - we are not considering helical waveguides such as those analysed by e.g. Treyss\`{e}de \cite{treyssede2008elastic}.

\begin{figure}[ht!]
    \centering
    \includegraphics[width = \textwidth]{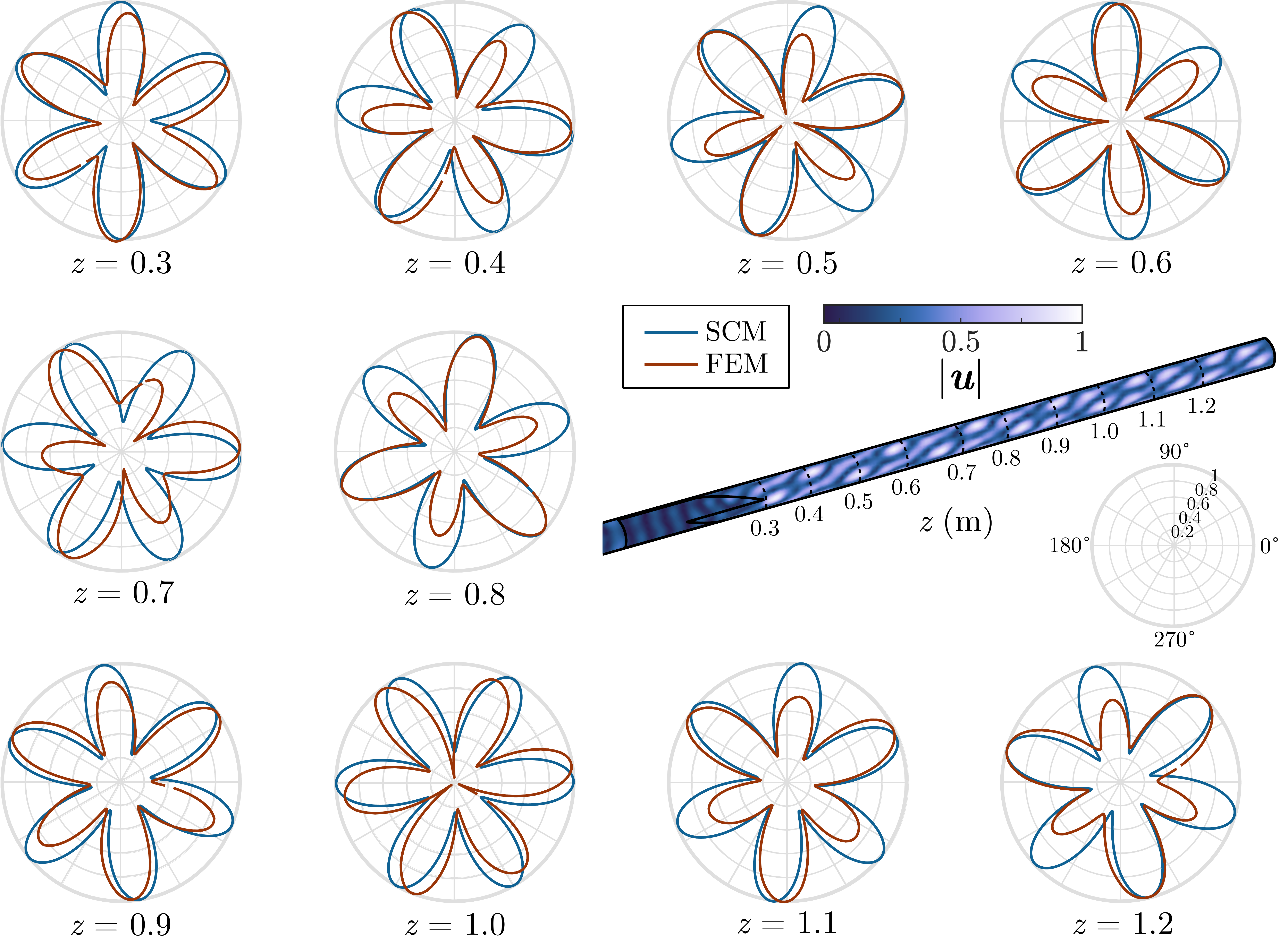}.
    \caption{\textbf{Constant Angular Profile:} Comparisons of angular profiles (normalised $|\bu|$) of the $F(3,2)$ mode calculated by a spectral collocation method (blue)  against those extracted from the FEM simulation in Figure~\ref{fig:hero}(a) (orange), at various distances along the pipe after the eSPP device. The field is taken on the outer surface of the pipe. Shown too is a zoom of the solid displacement component from Figure~\ref{fig:hero}(a).}
    \label{fig:hero2}
\end{figure} 

We highlight the uni-modal nature of the conversion in Figure~\ref{fig:hero2}, plotting the angular profiles of the normalised solid displacement around the outer pipe radius and at various distances along the length of the pipe, after the eSPP device. Additionally shown here are the mode shapes of a perfect $F(3,2)$ mode, reconstructed from the eigensolution of the spectral collocation method outlined in \ref{sec:Append1}. We emphasise here that the non-axisymmetric mode shapes produced after the device are the result of an axisymmetric excitation. Without the phase profile introduced by the eSPP these patterns are not producible with axisymmetric forcings due to the orthogonality of the mode shapes \cite{li2001excitation}. It can be seen here that the angular profile is constant along the length of the pipe, up to a modulation by the propagation factor $\mathrm{\exp}[\mathrm{i}(k_z z - \omega t)]$, i.e. the mode shape simply rotates around the pipe, where $k_z$ is the wavenumber along the axial direction of the pipe, $\omega$ the frequency of excitation and $t$ being time. Therefore unlike the complicated mode shapes produced by non-axisymmetric partial loading (e.g. Li and Rose \cite{li2001excitation}), here a predictable angular profile is present due to the single flexural $F(3,2)$ mode that has been excited due to the tailored mode conversion.

Throughout this paper we draw on the literature from the electromagnetic community to facilitate the design of the eSPP. Recent translations of optical phenomena to elastic devices have focused on concepts centred around metamaterials, gradient index (GRIN) lenses and topological insulators, and the discrete (often subwavelength) components that comprise them \cite{colombi2016seismic,colombi2017enhanced,chaplain2020tailored,chaplain2020topological,colombi2016resonant}. Some of these concepts have also recently been applied to slowing and focussing waves in pipe structures \cite{danawe2020conformal,Danawe2021}. Unlike these discrete, resonant devices we aim to leverage and emulate classical, continuous optical components - namely the optical spiral phase plate. In Section~\ref{sec:Sec2} we outline the conventional oSPP, including its history, uses and how its corresponding physics can be translated across to the elastic spiral phase pipe. Important differences between the analogous devices, that arise due from the nuances of elasticity, shall be presented and prove key to its design. We then introduce three possible configurations of the proposed eSPP devices, highlighting potential advantages and disadvantages of each. In Section~\ref{sec:Sec3} we choose the simplest of these configurations for experimental verification and proof of concept, detailing the design process and presenting the experimental confirmation of the passive mode conversion produced by the eSPP. We then finally draw conclusions and outline future perspectives and applications for the eSPP.




\section{Optical Spiral Phase Plates and Elastic Spiral Phase Pipes}
\label{sec:Sec2}
The first optical spiral phase plates, or spiral-phase-delay plates, were realised in the early 1990s \cite{LaserFocusWorld}, enabling light fields with an azimuthally varying phase profile to be created. These laser profiles enabled the probing of optical vortices \cite{arecchi1991vortices}, have uses in particle acceleration experiments \cite{tidwell1993efficient}, and crucially enabled investigations into angular momentum and spin-orbit coupling of photons \cite{kristensen1994angular,beijersbergen1994helical}. Spiral phase plates can be used to generate so called Laguerre-Gaussian (LG) modes, helical-wavefront beams, or ``doughnut" shaped beams of light that satisfy the paraxial wave equation which, after Allen et al's realisation that these carry optical angular momentum \cite{allen1992orbital}, and the subsequent experimentation \cite{o2002intrinsic}, re-ignited interest in the field of optical tweezers \cite{barnett2017optical,he1995direct,garces2003observation}. The helical phase profiles generated by the humble spiral phase plate have inspired many analogues across a variety of wave systems \cite{wunenburger2015acoustic,jin2019flat,fan2020acoustic}.

Despite the prevalence of the oSPP, and the designs it has inspired, there has been, until now, no elastic equivalent. This is perhaps surprising, especially in pipe structures which already posses the topological features necessary to support helical phased wave fronts. Below we outline the original oSPP and then translate this to form the eSPP; nuances in the elastic system due to the dispersive nature of the elastic waves in pipes require additional thought in its design.

\begin{figure}
    \centering
    \includegraphics[width = 0.775\textwidth]{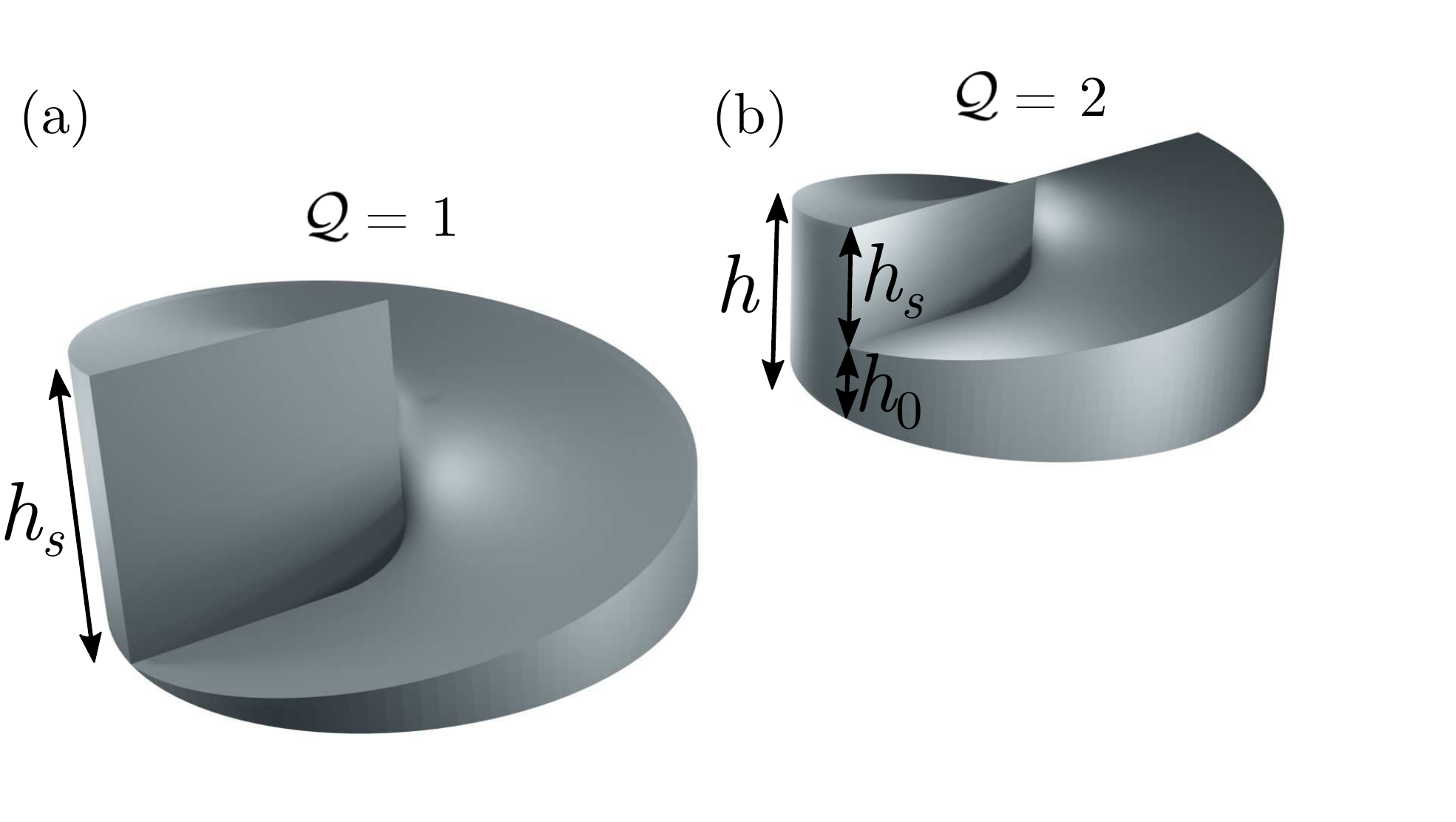}
    \caption{\textbf{Optical spiral phase plates:} Classical designs capable of producing LG beams with example topological charges $\mathcal{Q} = 1, 2$ respectively. The step height $h_s$ is shown, such that the total extent of the device is $h = h_s + h_0$.}
    \label{fig:oSPP}
\end{figure}

Figure~\ref{fig:oSPP}(a) shows the classical optical spiral phase plate. It consists of a plate whose thickness varies proportionately to the azimuthal angle $\theta$ defined about the central middle point of the plate and is thus parameterised as a circular helicoid. The helical surface resembles one turn of a staircase \cite{beijersbergen1994helical}, the height of which we denote $h_s$. Figure~\ref{fig:oSPP}(b) shows another configuration, which includes a region of uniform thickness, $h_0$, before the spiral region. The total height at any given point on the spiral is then given as $h = h_s + h_0$. Spiral phase plates are used to produce optical vortex beams, characterised by a phase singularity at the beam centre with locally vanishing intensity carrying a topological charge defined as \cite{oemrawsingh2004production}
\begin{equation}
    \mathcal{Q} = \frac{1}{2\pi}\oint d\chi,
\end{equation}
where $\chi$ is the phase of the field. This quantity locally characterises the complex field amplitude, $u$ through
\begin{equation}
    u(r,\theta,z) = u^{\prime}(r,z)\mathrm{\exp (\mathrm{i}\mathcal{Q}\theta)},
\end{equation}
where $u^{\prime}$ is a complex valued function with $r$ and $z$ the radial and axial directions, with propagation along $z$. 

When inserted, for example in the waist of a Gaussian beam, the oSPP endows the incident field with an azimuthally dependent phase retardation. In optics this phase profile is given by the difference in optical path length experienced by the beam as it traverses the plate, i.e. 
\begin{equation}
    \varphi(\theta,\lambda) = \frac{2\pi}{\lambda}\zeta,
\end{equation}
where $\zeta$ is the optical path difference, i.e. the different lengths `seen' by the wave due to the change in phase speed by refraction. This yields the phase profile
\begin{equation}
    \varphi(\theta,\lambda) = \frac{2\pi}{\lambda}\left[\frac{(\Delta n) h_s\theta}{2\pi} + nh_0 \right],
\end{equation}
where $\Delta n = (n - n_0)$, with $n$ the refractive index of the plate and $n_0$, $\lambda$ being the refractive index of, and wavelength in, the surrounding medium respectively. Given a uniform phase change is imposed throughout the initial height $h_0$. the vortex charge $\mathcal{Q}$ is then given as \cite{oemrawsingh2004production}
\begin{equation}
    \mathcal{Q} = \frac{(n - n_0)h_s}{\lambda}.
    \label{eq:Q}
\end{equation}
The sign of $\mathcal{Q}$ determining the handedness of the helical phase-front. We adopt the convention of $\mathcal{Q} > 0$ being left-handed and $\mathcal{Q} < 0$ being right-handed. Alternatively the step height can be partitioned equally across $\mathcal{Q}$ number of turns \cite{cano2018dynamic} as we show in Figure~\ref{fig:oSPP}(b). This simple equation then allows the step height to be determined to generate a particular topological charge for a given wavelength, displaying the chromatic nature of the oSPP. In optics, generation of low $\mathcal{Q}$ values requires either the step height to be commensurate to the order of the wavelength or for the surroundings to be almost index matched such that $\Delta n \approx 0$ \cite{oemrawsingh2004production,beijersbergen1994helical}. 

\begin{figure}
    \centering
    \includegraphics[width = \textwidth]{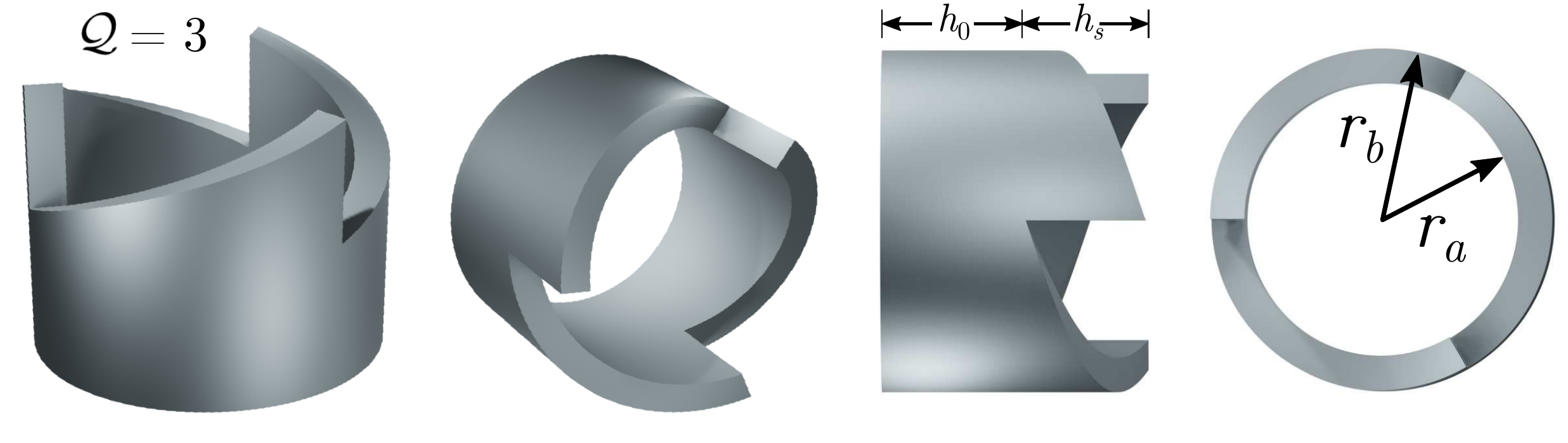}
    \caption{\textbf{An Elastic Spiral Phase Pipe:} Schematics of a three-step elastic spiral phase pipe of inner radius $r_a$ and outer radius $r_b$.}
    \label{fig:eSPP}
\end{figure}

We now turn our attention to the elastic analogue of the oSPP, and outline the similar design for the elastic spiral phase pipe. Figure~\ref{fig:eSPP} outlines an example eSPP structure, which in essence is an elastic, hollowed spiral phase plate. Here we show a pipe of inner radius $r_a$ and $r_b$ with three spirals. Due to the coupled nature of the elastic wave system (\ref{sec:Append1}) endowing an incident wave with, say a helical phase profile, does not merely change the amplitude profile of the wave, but couples the motion into, potentially, all three coordinate directions (i.e. excites $u_r$, $u_\theta$ and $u_z$). Therefore in the elastic system it is not appropriate to just consider the phase change of one component, but necessary to consider mode conversion from one wave type to another. We shall demonstrate that the eSPP is capable of mode conversion from a given incident compressional wave to a desired flexural wave.  

\begin{figure}
    \centering
    \includegraphics[width = 0.85\textwidth]{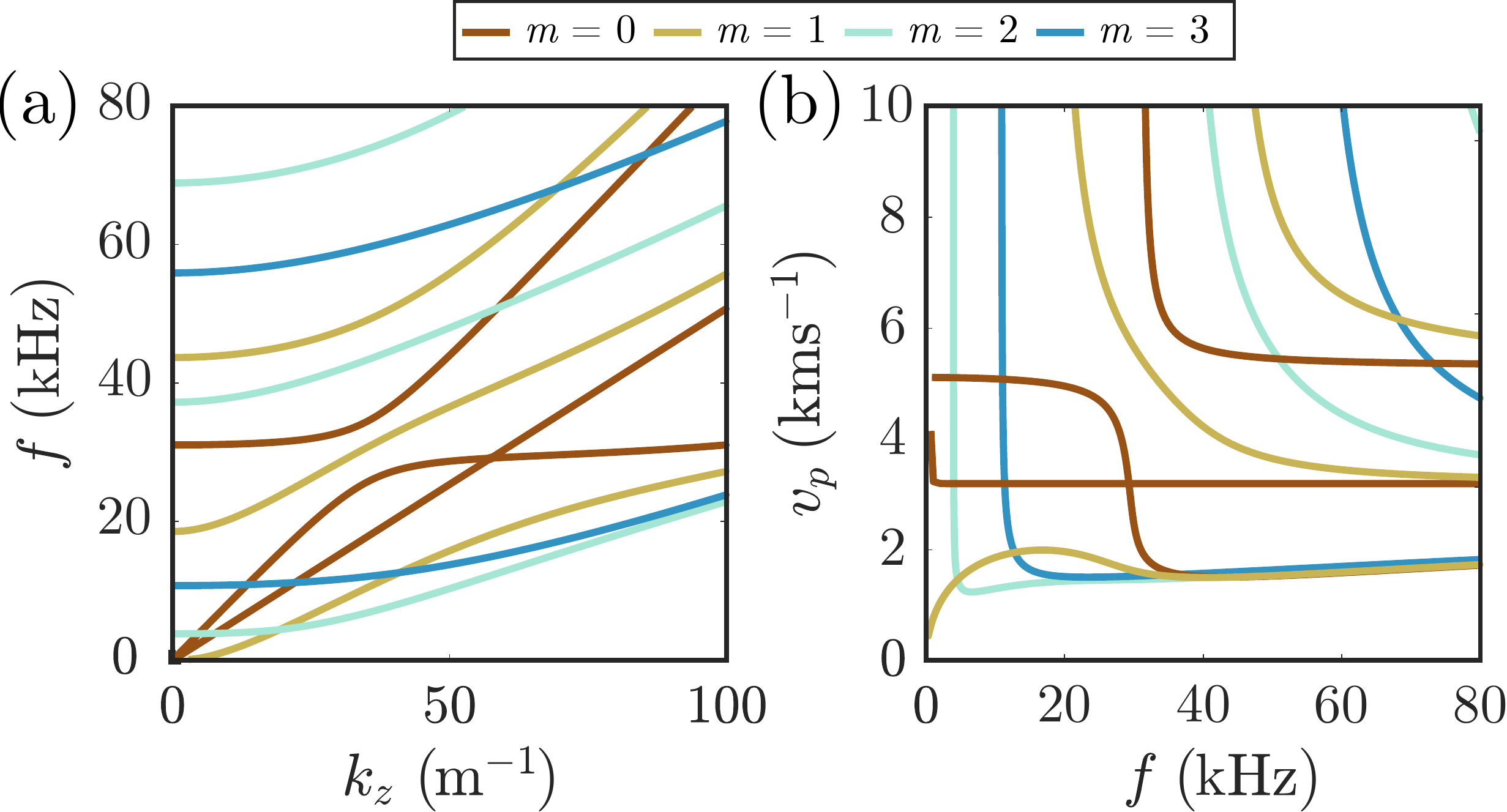}
    \caption{\textbf{Dispersion curves of a steel pipe:} (a) Conventional representation in physics, frequency vs wavenumber (b) Conventional representation in engineering, phase velocity vs frequency. Evaluated by the spectral collocation method (SCM) shown in \ref{sec:Append1}, for inner diameter $5~\si{\centi\meter}$ and thickness $4.5~\si{\milli\meter}$.}
    \label{fig:disp}
\end{figure}

Fortunately, in the elastic systems we shall consider, having step heights of comparable size to the incident wavelength is straightforward. However, the analogy to \eqref{eq:Q} is more nuanced in elasticity due to the lack of the notion of a refractive index; in optical systems the refractive index is defined as the ratio of the phase speeds of light in vacuo compared to that in a given material, i.e. $n = c/v$. For most materials in electromagnetism, the dispersionless nature of the supported waves and the universal speed limit of $c$ then uniquely defines this refractive index. In elastic materials however, even those that are isotropic and homogeneous, there exists distinct wave types (e.g. compression and shear) with  unique phase speeds and that couple at interfaces \cite{graff2012wave}. This is clearly seen by the dispersion curves for the first four circumferential orders ($m = 0,\ldots,3$) in Figure~\ref{fig:disp}, where we show two conventional representations used in physics and engineering. 

We account for the lack of definiteness in the definition of refractive index in elastic systems by defining a relative refractive index, relating the phase speeds of a given incident mode, $c_i$ (with wavelength $\lambda_i = 2\pi/k_i$), to the desired flexural mode with a helical-phase front and phase speed $c_f$. We therefore define this relative refractive index $\tilde{n} = c_f/c_i$. Continuing with this we arrive at the following formula for the step profile of the eSPP
\begin{equation}
    h_s = \frac{2\pi\mathcal{Q}}{k_i(\tilde{n}-1)}. 
    \label{eq:hs}
\end{equation}

The simple derivation from the optical analogue highlights another nuance with the elastic system. Rearranging \eqref{eq:Q} we see that $h_s = \lambda|\mathcal{Q}(n-n_0)|$, and as such is unchanged to matter which direction the incident wave traverses the spiral phase plate. However for the eSPP, due to the form of $\tilde{n}$, reciprocity of the mode conversion effects in the eSPP is not guaranteed - it is dependent on the form of the incident wave upon the spiral region.

\begin{figure}
    \centering
    \includegraphics[width = \textwidth]{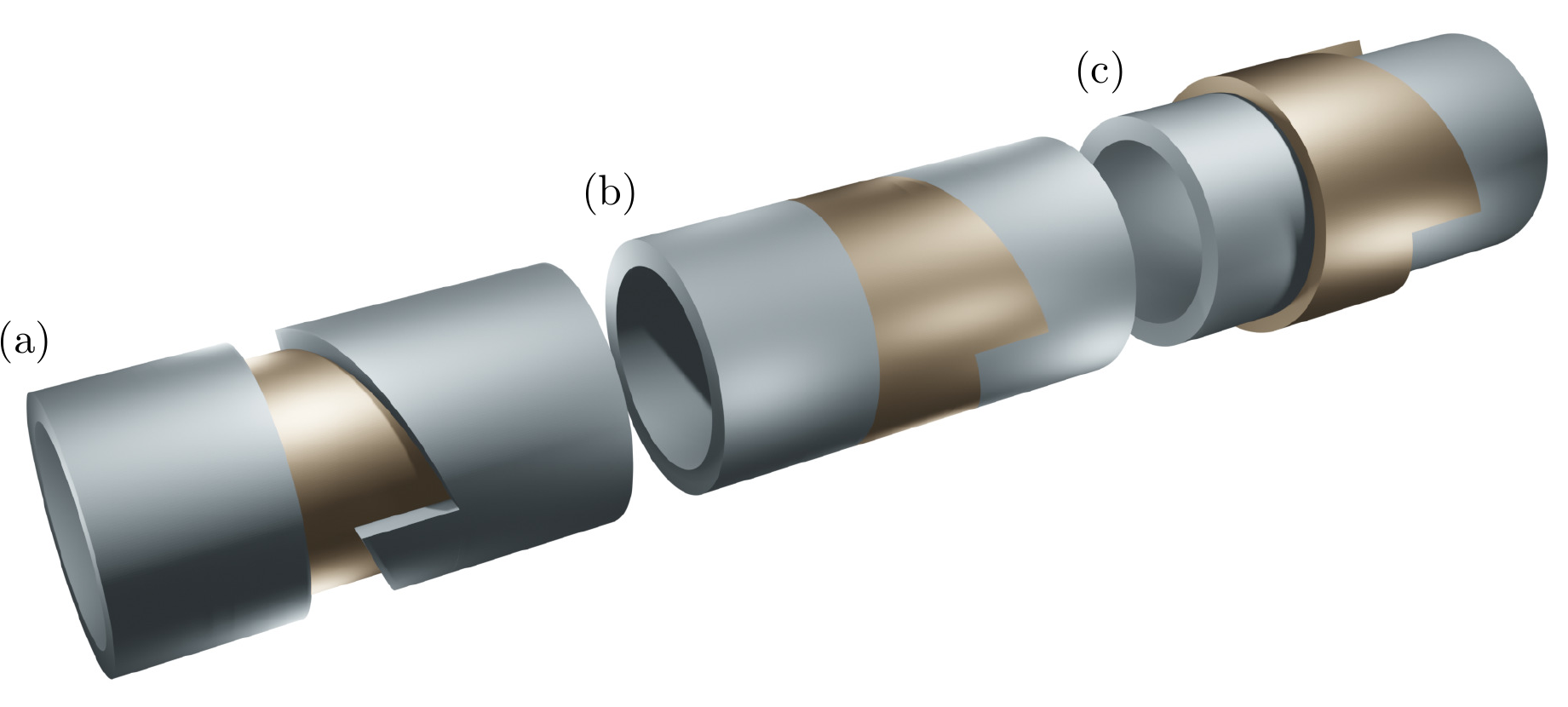}
    \caption{\textbf{eSPP configurations:} (a) Notch (b) Join and (c) Collar; the bronzed regions represent the spiral phase pipe.}
    \label{fig:configs}
\end{figure}

Given that the ultrasonic waves we consider are guided along the length of the pipe we show, in Figure~\ref{fig:configs}, three variations on the design of the eSPP. Figure~\ref{fig:configs}(a) shows a notch configuration whereby a portion of the pipe is milled out, leaving an impression of the eSPP (bronzed region in Figure~\ref{fig:configs}(a)). This then delineates the pipe into two regions with different thicknesses. We define these as $h_1$ and $h_2$ such that $h_1 < h_2$. The profile of the eSPP is chosen such that the $L(0,2)$ mode in the eSPP (thickness $h_1$) mode converts to the $F(3,2)$ mode (or whichever desired mode) in the thicker region of thickness $h_2$. The advantages of this is that the region comprising the eSPP is the same material of the pipe. Disadvantages are that the pipe has to be machined, and that the thickness $h_1$ affects the coupling of the incident $L(0,2)$ wave to the desired converted mode. 

Figure~\ref{fig:configs}(b) shows the join configuration, used in Figures~\ref{fig:hero}\&\ref{fig:hero2}, in which an eSPP of different material is inserted between two regions of a pipe. The advantages of this is that the entirety of the $L(0,2)$ mode is incident on the face of the eSPP and so has the greatest conversion efficiency. Additionally the flexibility of having an eSPP of different material at the join enables a wider range of dispersive properties to be designed. Again drawbacks from milling are now present in two pipes and joining them with either welds or adhesives is required.

The final design is shown in Figure~\ref{fig:configs}(c) and is termed the collar configuration. This can be thought of as an inverted notch, whereby either the whole pipe, apart from the eSPP region, is milled (leaving $h_1 > h_2$). However a more practical design is to have a separate collar which fits or fastens around the pipe, possibly in several components. Again this has the advantage that it can be a different material from the pipe, and be removed after its used; its conversion efficiency is however dependent on the contact between the pipe and the eSPP.

For simplicity we opt to verify the operation of the eSPP with the notch configuration. In Section~\ref{sec:Sec3} we highlight its design and validate the conversion from $L(0,2)$ into $F(3,2)$ in an aluminium pipe.

\section{\label{sec:Sec3} Experimental Verification}

\subsection{Design}
In order to design a suitable eSPP for any given mode conversion the effective refractive index between the two wave speeds must be evaluated. As such, the dispersion curves for each region of the pipe (i.e. the eSPP and the rest of the pipe) are required. We evaluate these for an infinitely long region of each pipe by employing a spectral collocation method (\ref{sec:Append1}). Once obtained, the dispersion curves are then used to infer the supported modes in each section of the pipe. We first consider an infinitely long aluminium pipe, of inner diameter $40~\si{\milli\meter}$ and thickness $h_2 = 10~\si{\milli\meter}$ of density $\rho = 2710~\si{\kilo\gram\meter^{-3}}$, Young's Modulus $E = 70~\si{\giga\pascal}$ and Poisson's ratio $\nu = 0.33$. We denote the thickness of this region $h_2$; this is to be the section of the pipe where the desired $F(3,2)$ mode to be excited, after conversion by the eSPP. We then design an eSPP region by considering an aluminium pipe with the same internal diameter but now with $h_1 = 4~\si{\milli\meter}$. The thickness of the eSPP is chosen such that $h_1 < h_2/2$ to maximise the coupling to the spiral region. The infinite dispersion curves for the eSPP region are then calculated, and at a desired frequency the relevant step height can be evaluated through Eq.~\eqref{eq:hs}.

\begin{figure}[ht!]
    \centering
    \includegraphics[width = \textwidth]{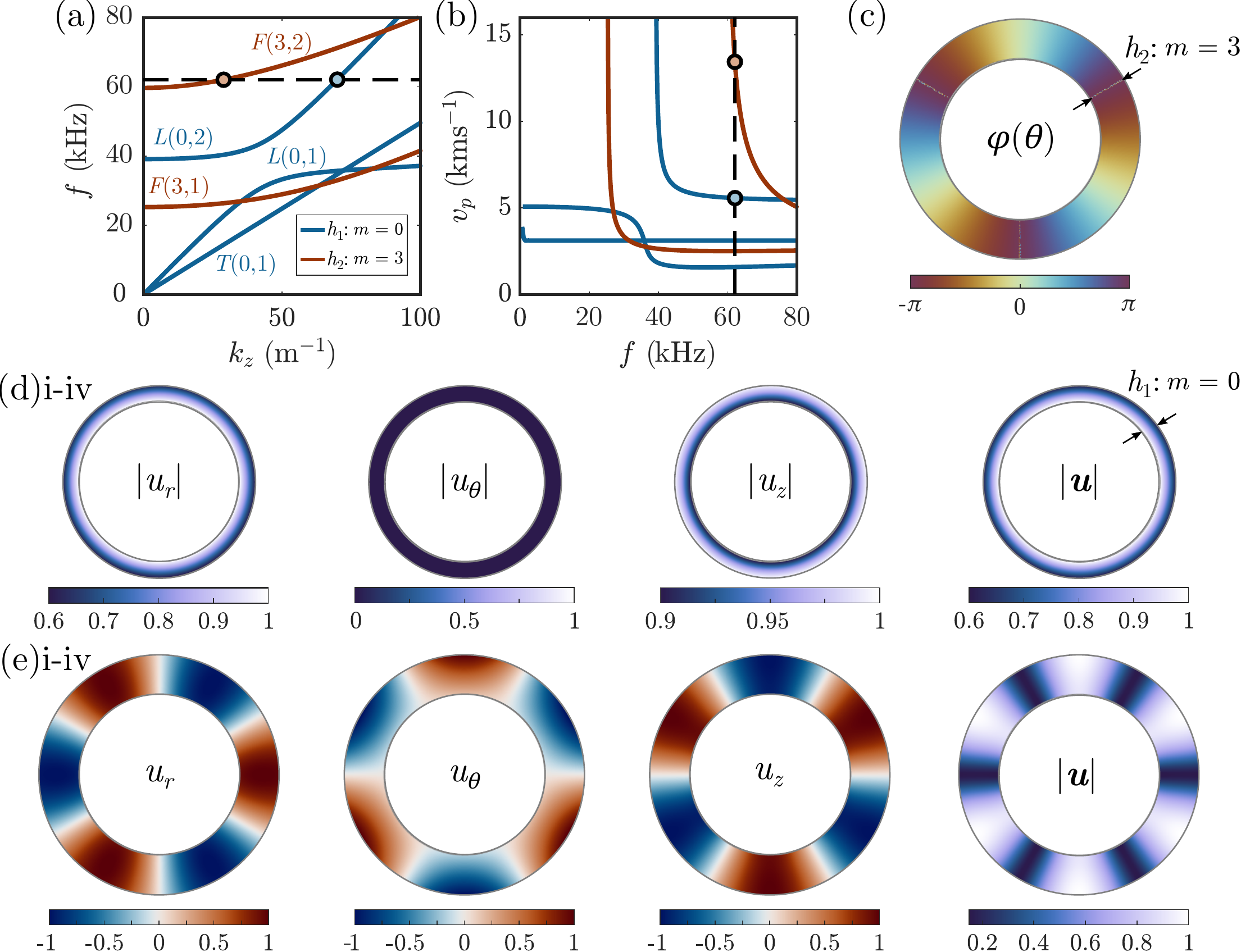}
    \caption{\textbf{Dispersion Design and Mode Shapes:} (a) Physicists representation of dispersion curves (b) Engineers representation of dispersion curves. (c) Phase profile of $\varphi(\theta)$ in region of thickness $h_2$, confirming $\mathcal{Q} = 3$. (di-iv) are the normalised radial, angular, axial and total solid displacement fields respectively in $h_1$. (ei-iv) show the same fields but in $h_2$. These fields are reconstructed from the SCM eigensolutions (\ref{sec:Append1}).}
    \label{fig:Design}
\end{figure}

Figures~\ref{fig:Design}(a,b) show the $m = 3$ dispersion curves for the pipe of thickness $h_2$ and the $m = 0$ dispersion curves for the eSPP region with thickness $h_1$. We wish to excite the $L(0,2)$ in the eSPP to mode convert this to the $F(3,2)$ mode in the thicker portion of the pipe, and so operate above $60~\si{\kilo\hertz}$ where the frequency spectra of the two families of modes overlap. We select, arbitrarily, a design frequency of $62~\si{\kilo\hertz}$ for the conversion. The phase speeds for each mode are then extracted from the dispersion curves in Figure~\ref{fig:Design}(b), and used to design the step height for the eSPP, determined to be $h_s = 190~\si{\milli\meter}$. Figure~\ref{fig:Design}(c) shows the angular phase of the $F(3,2)$ mode, confirming it indeed has topological charge $\mathcal{Q} = 3$, as expected. In Figures~\ref{fig:Design}(d)i-iv and Figures~\ref{fig:Design}(e)i-iv we show the mode shapes for the longitudinal $L(0,2)$ mode in the eSPP region and the $F(3,2)$ mode in the thick pipe, showing that they are indeed orthogonal and so require the inhomogeneity introduced by the eSPP to couple. 


\subsection{Experimental Verification}
\begin{figure}[t!]
    \centering
    \includegraphics[width = \textwidth]{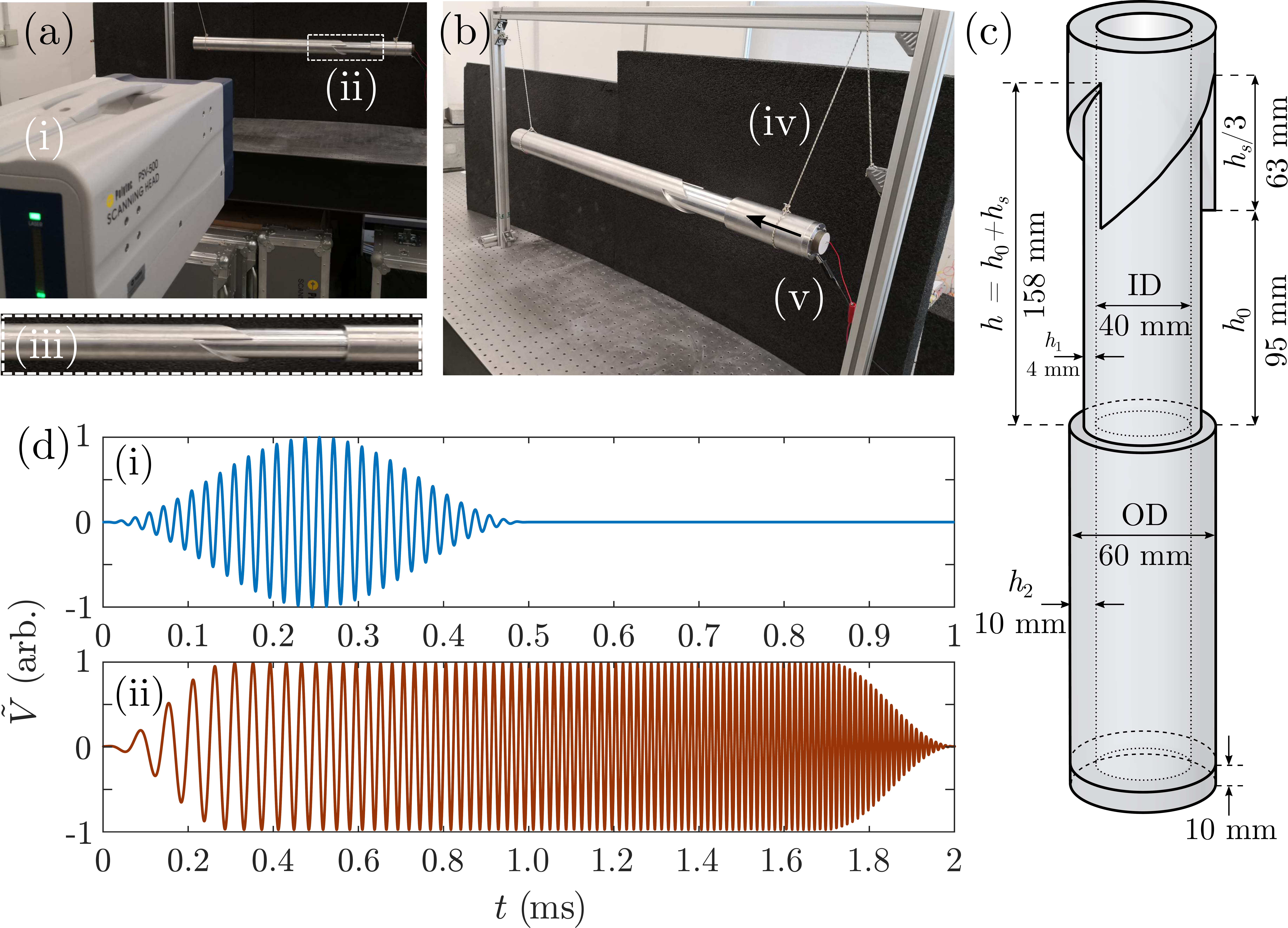}
    \caption{\textbf{Experimental Setup:} (a) Measurements are performed using a 3D laser vibrometer (i) scanning the velocity field on one side of the prototype (ii) made of a pipe with a notched eSPP configuration (iii). (b) The pipe is suspended through elastic cables on a frame in order to avoid edge disturbances (iv) while on the right side a piezolectric disk is glued to the cap in order to provide excitation of longitudinal waves (v). (c) Shows schematic detailing geometric parameters. (d) Time domain source of piezo-disk for experimental results in Figure~\ref{fig:Snapshots} (blue) and sweep source for the experimental Fourier spectra in Figure~\ref{fig:numFFT} (orange).}
    \label{fig:Setup}
\end{figure}

To validate our findings, we perform time domain experiments, and compared to numerical simulations, on an aluminium pipe 
where the eSPP is CNC milled out removing the material from the thickness. This ensures a region of $h_2$ before and after the eSPP of thickness $h_1$. The pipe is $900~\si{\milli\meter}$ long, with internal and external diameter equal to $40~\si{\milli\meter}$ and $60~\si{\milli\meter}$ respectively. The eSPP is obtained by milling over a thickness of $6~\si{\milli\meter}$. To decrease the length of the step height, $h_s$, we partition it over three steps, rather than a single step. We also include a length of $h_0$ to ensure the $L(0,2)$ mode is excited in the eSPP. The total portion of the pipe of length is then of $h = h_0 + h_s/3 = 158~\si{\milli\meter}$. The experimental setup is shown in Figure~\ref{fig:Setup}, where the pipe is suspended through elastic cables in order to avoid disturbances from the boundaries. At the right boundary, an aluminium disk of radius $60~\si{\milli\meter}$ and thickness $10~\si{\milli\meter}$ is connected through a set of screws to the pipe, while a  piezoelectric disk with diameter $35~\si{\milli\meter}$ and thickness $12~\si{\milli\meter}$ is glued to provide excitation.

\begin{figure}[t!]
    \centering
    \includegraphics[width = \textwidth]{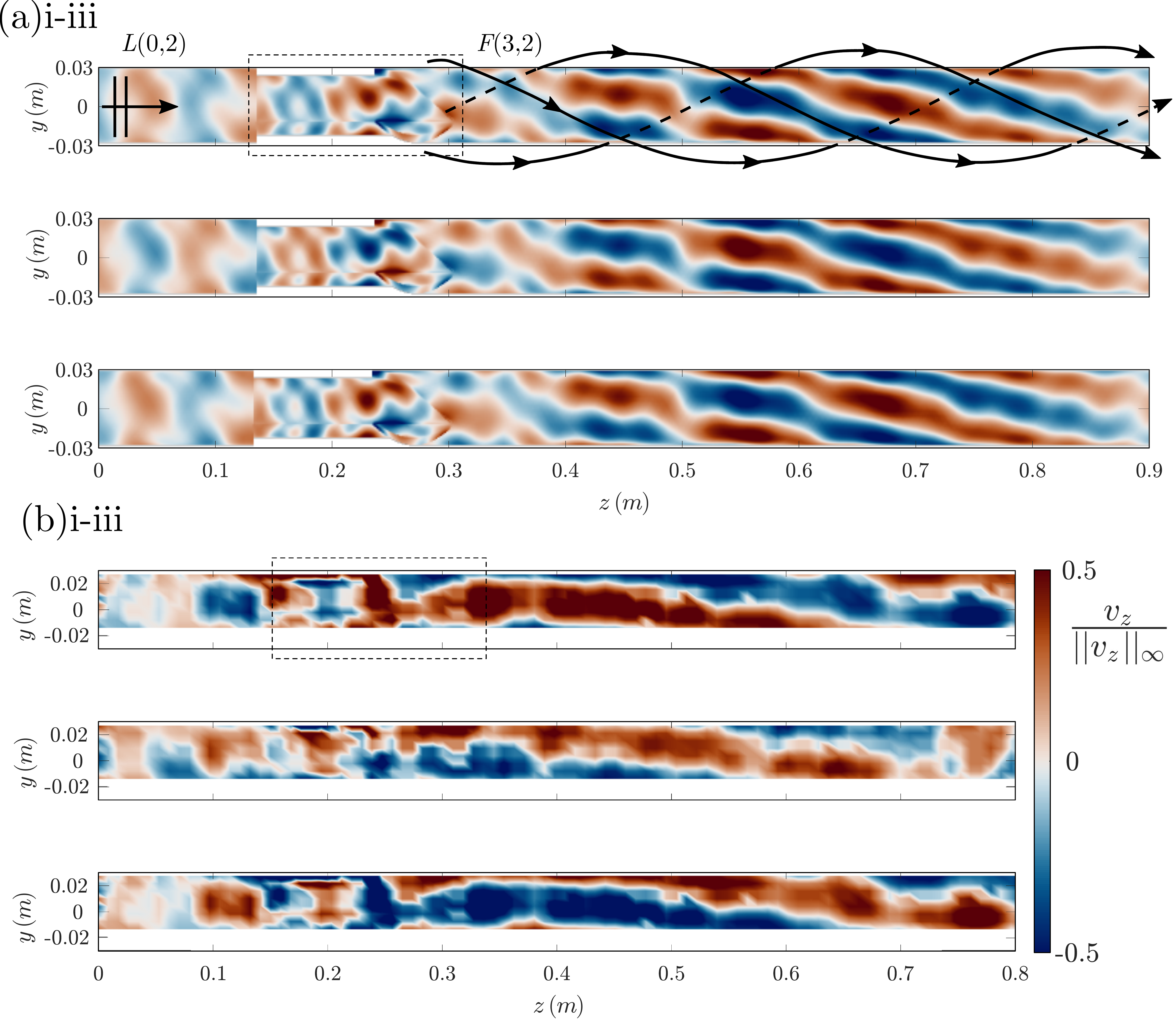}
    \caption{\textbf{Time Domain Results:} Snapshots in time of the normalised velocity field $v_z$ for (a)(i-iii) Numerical simulations at $t = 0.812$, $0.82$ and $0.828~\si{\milli\second}$ respectively and (b)(i-iii) Experimental results  at $1.248$, $1.256$ and $1.264~\si{\milli\second}$ respectively. Mode conversion of longitudinal $L(0,2)$ waves into flexural $F(3,2)$ waves after the eSPP (highlighted in dashed rectangle) is clearly seen. We note the handedness of the helical phase-fronts differ compared to that in e.g. Figure~\ref{fig:hero} as we measure the field right to left in the experiments. The input signal for both cases is as in Figure~\ref{fig:Setup}(d)(i). Due to the curvature of the pipe we can only extract the displacement fields over the portion shown in the experimental results, compared to the whole pipe from the simulations (note the different axes).} 
    \label{fig:Snapshots}
\end{figure}

The wavefield is measured on one side of the pipe through a Polytec 3D Scanner Laser Doppler Vibrometer (SLDV), which is able to separate the out-of-plane velocity field in both space and time. A narrow-band spectrum excitation of central frequency $f_c=60~\si{\kilo\hertz}$ and fractional bandwidth $B=\Delta f/ f_c = 0.14$ is synchronously started with the acquisition which, in turn, is averaged in time to decrease the noise. Figure~\ref{fig:Snapshots} compares the numerical and experimental wavefield along the pipe at different time instants. The agreement between the numerical and the experimental data confirms the capability of the eSPP to generate an helical phase profile, mode converting incoming longitudinal $L(0,2)$ waves into flexural $F(3,2)$ waves.

\begin{figure}[t!]
    \centering
    \includegraphics[width = 1\textwidth]{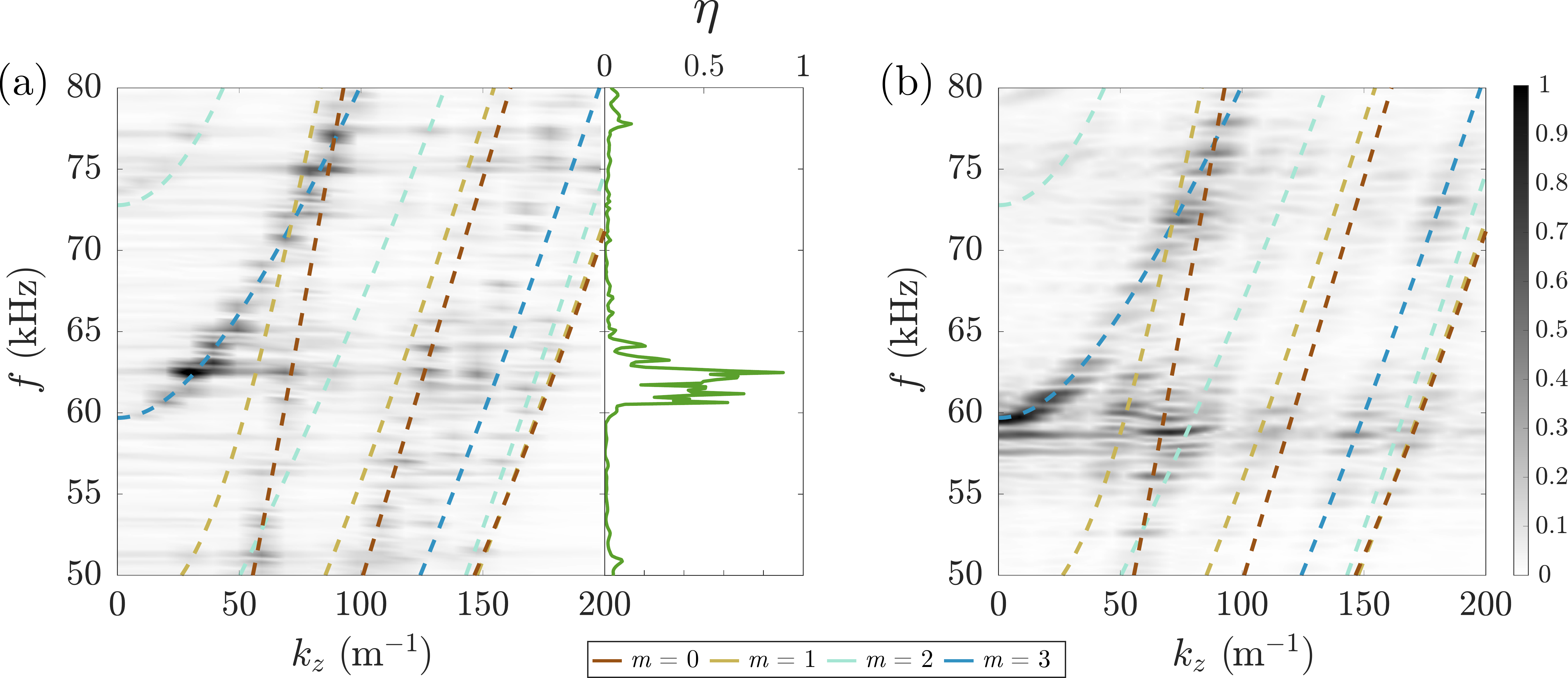}
    \caption{\textbf{Normalised Fourier Spectra and conversion efficiency:} (a) FFT of FEM time domain simulation performed with COMSOL MultiPhysics\textsuperscript{\textregistered} \cite{comsolSolidMech}. The peak at the design frequency of $62~\si{\kilo\hertz}$ is clear on the $F(3,2)$ branch. This is highlighted by the predicted conversion efficiency, $\eta$ (right panel). (b) FFT of Experimental data. In both cases the frequency sweep source (Figure~\ref{fig:Setup}(dii)) is used and the FFT performed along a line after the eSPP on the surface of the pipe. Clear excitation of the $F(3,2)$ mode is seen in both cases.}
    \label{fig:numFFT}
\end{figure}
To assess the degree of conversion we inspect and compare the Fourier spectra obtained by the spatio-temporal Fast Fourier Transform (FFT) along a line after the eSPP, shown in Figure~\ref{fig:numFFT}, for the time domain FEM simulation (a) and the experimental results (b). Given the finite length of the pipe we are limit in the resolution in $k_z$, and so we additionally show, in \ref{sec:appendB}, the results of the FFT for a frequency domain simulation over a much longer pipe that is $10~\si{\meter}$ in length. Overlaid on these spectra are the dispersion curves for the infinitely long aluminium pipe of thickness $h_2$. Clear agreement is seen, and that the eSPP is most efficient near the design frequency. We elucidate this further by showing, in the side panel of Figure~\ref{fig:numFFT}(a), a measure of the conversion efficiency, $\eta$. We evaluate this from a frequency domain FEM simulation, and define $\eta$ as the ratio of the integrals of the absolute values of the compressional and rotational complex mechanical energy flux, over the same volume before and after the eSPP, such that 
\begin{equation}
    \eta = \frac{\int|\boldsymbol{F}_z|^2dV}{\int|\boldsymbol{F}_\theta|^2dV'},
\end{equation}
where $\boldsymbol{F} = \sigma\cdotp\boldsymbol{u}^{*}$, with $\sigma$ the stress tensor and $\boldsymbol{u}^{*}$ the complex conjugate of the solid displacement field. We denote the volume element before and after the eSPP as $dV$ and $dV'$ respectively, ensuring that the total integral volume is the same in each case. This definition of the conversion efficiency then gives a measure of the energy present in the compressional mode before the eSPP and the flexural mode afterwards in the rest of the pipe. As seen in Figure~\ref{fig:numFFT}(a) we achieve a theoretical conversion efficiency of $> 90\%$ at the design frequency. We note that this calculation neglects any damping effects.

\section{Conclusions}

We have presented an elastic analogue to the optical spiral phase plate - the elastic spiral phase pipe - and shown that it can be used to efficiently mode convert axisymmetric longitudinal guided modes in pipes into arbitrary non-axisymmetric flexural modes. We achieved this by defining the relative refractive index, which relates the phase velocities of the two considered modes to the azimuthally varying step profile of the device. Moreover we presented three possible configurations of this device, each with their own advantages and disadvantages. This was enabled by obtaining the dispersion relations of the two sections of the pipe through a spectral collocation method. We demonstrated, through numerical simulation corroborated by experimentation that the $L(0,2)$ mode can be efficiently converted to the $F(3,2)$ mode at the design frequency, chosen here to be $62~\si{\kilo\hertz}$. 

Advantages of this new device include that, compared to more complex non-axisymmetric forcings, an axisymmetric forcing can be used to generate a flexural wave with constant angular profile. As such we envisage exciting applications for this device in the fields of non-destructive testing, ultrasonic motor design and beyond.
 
\section*{Acknowledgments}
The authors are grateful to Prof. Richard V. Craster, Prof. Michael J.S. Lowe, and Evan Hill Esq. for helpful conversations.
G.J.C gratefully acknowledges financial support from the EPSRC in the form of a Doctoral
Prize Fellowship, and from the Royal Commission for the Exhibition of 1851 in the form of a Research Fellowship.
J.M.D.P acknowledges the financial support from the  H2020 FET-proactive  project MetaVEH under grant agreement No. 952039.

\appendix
\section{\label{sec:Append1} Spectral Collocation Method}


Guided propagating waves in a hollow isotropic elastic cylinder are governed by Navier's equation
\begin{equation}
    \mu\nabla^2\bu + (\lambda + \mu)\nabla(\nabla\cdotp\bu) = \rho\ddot{\bu},
\end{equation}
where $\bu$ is the displacement field, $\lambda$ and $\mu$ are Lam\'{e}'s first and second parameters respectively, $\rho$ the material density and $\ddot{\bu}$ is the acceleration. As the material is isotropic, Helmholtz decomposition is used to write the displacement in terms of the dilatational scalar potential, $\Phi$ and the equivoluminal vector potential $\bPsi$
\begin{equation}
    \bu = \nabla\Phi + \nabla\times\bPsi.
\end{equation}
Under this assumption, Navier's governing wave equation then reduces to two wave equations for compressional and shear waves:
\begin{equation}
\begin{cases}
    \nabla^2\Phi = c_p^{-2}\ddot{\Phi}, & c_p = \sqrt{\frac{\lambda + 2\mu}{\rho}} \\
    \nabla^2\bPsi = c_s^{-2}\ddot{\bPsi}, & c_s = \sqrt{\frac{\mu}{\rho}},
\end{cases}
\label{eq:waveeqns}
\end{equation}
with $c_p$ and $c_s$ being the compressional and shear bulk wavespeeds respectively. Along with these two equations we employ traction free boundary conditions on the inner and outer radii, $r_a$ and $r_b$, such that
\begin{equation}
    \sigma_{rr} = \sigma_{r\theta} = \sigma_{rz} = 0\big\rvert_{r_{a,b}},
    \label{eq:tractionfree}
\end{equation}
along with the infinitely long cylinder gauge invariance through
\begin{equation}
    \nabla\cdotp\Psi = 0.
    \label{eq:gauge}
\end{equation}
Working in cylindrical coordinates, such that $\bPsi = (\Psi_{r},\Psi_{\theta},\Psi_{z})$, then gives
\begin{equation}
\begin{split}
    \nabla^2\Phi &= \left(\frac{\mathrm{\partial}^2}{\mathrm{\partial} r^2} +\frac{1}{r}\frac{\mathrm{\partial}}{\mathrm{\partial} r} + \frac{1}{r^2}\frac{\mathrm{\partial}^2}{\mathrm{\partial}\theta^2} + \frac{\mathrm{\partial}^2}{\mathrm{\partial} z^2}\right)\Phi, \\ \\ 
    \nabla^2\bPsi &= \nabla^2\left(\Psi_{r}\boldsymbol{\hat{e}}_{r} + \Psi_{\theta}\boldsymbol{\hat{e}}_{\theta} + \Psi_{z}\boldsymbol{\hat{e}}_{z}\right) \\
    &= \left(\nabla^2\Psi_{r} - \frac{1}{r^2}\Psi_{r} -\frac{2}{r^2}\frac{\mathrm{\partial}\Psi_{\theta}}{\mathrm{\partial}\theta}\right)\boldsymbol{\hat{e}}_{r} \\&+ \left(\nabla^2\Psi_{\theta} - \frac{1}{r^2}\Psi_{\theta} +\frac{2}{r^2}\frac{\mathrm{\partial}\Psi_{r}}{\mathrm{\partial}\theta} \right)\boldsymbol{\hat{e}}_{\theta} \\&+ \left(\nabla^2\Psi_{z}\right)\boldsymbol{\hat{e}}_{z}.
\end{split}
\end{equation}
Following Rose \cite{rose2014ultrasonic} and Gazis \cite{gazis1959a,gazis1959b} we pose the complete solutions for guided axial waves
\begin{equation}
\begin{split}
    \Phi &= \phi(r)\mathrm{\exp}[\mathrm{i}(m\theta + k_{z}z - \omega t)], \\
    \Psi_{r} &= \psi_{r}(r)\mathrm{\exp}[\mathrm{i}(m\theta + k_{z}z - \omega t)], \\
    \Psi_{\theta} &= \psi_{\theta}(r)\mathrm{\exp}[\mathrm{i}(m\theta + k_{z}z - \omega t)], \\
    \Psi_{z} &= \psi_{z}(r)\mathrm{\exp}[\mathrm{i}(m\theta + k_{z}z - \omega t)], \\
\end{split}
\label{eq:Expansions}
\end{equation}
such that $m \in \mathbb{Z}$. Noninteger values of $m$ correspond to circumferential waves propagating in an isolated section of an annulus  \cite{adamou2004spectral}. Other cases for the solutions of circumferential waves include circumferential shear horizontal waves and circumferential Lamb-type waves \cite{rose2014ultrasonic}. In the case of these waves any displacement in the $z$-direction must be uniform through the entire $z$-plane, and it is necessary to distinguish between the circular and angular wavenumber \cite{viktrov1967rayleigh}.

At this stage we diverge from the methods suggested by Rose \cite{rose2014ultrasonic}, where substituting \eqref{eq:Expansions} into \eqref{eq:waveeqns} leads to four forms of Bessel's equation for the radial components of each potential. Solving for the dispersion relation then requires incorporating the boundary conditions \eqref{eq:tractionfree}\&\eqref{eq:gauge} that leads to an exercise in root-finding for the zeros of the determinant of the resulting system of equations \cite{rose2014ultrasonic,kumar1972dispersion}. This method has been used with success for multilayer and viscoelastic problems \cite{barshinger2004guided}, yet is tedious and particularly perilous if non-dimensionalisation is not used which leads to large arguments in the Bessel functions. Alternative methods also include the Finite Element Method \cite{manconi2009wave} (FEM) and Semi-Analytical FEM \cite{hayashi2003guided,mu2008guided} (SAFEM).

Instead we opt to expand the Spectral Collocation Method (SCM) of Adamou and Craster \cite{adamou2004spectral} which solves for the circumferential modes of an infinite cylinder. This method is especially attractive given its spectral accuracy, speed and ease of implementation. SCMs have had much success and been applied to many other scenarios surrounding pipes and pipe structures, and have been adapted to include viscoelasticity \cite{quintanilla2015guided}, anisotropy \cite{quintanilla2015modeling}, and fluid filled interactions \cite{karpfinger2008modeling}. The success of this method is highlighted by the resulting commercialisation \cite{pavlakovic1997disperse}.

The basic idea is to represent \eqref{eq:waveeqns} in operator form and construct a generalised eigenvalue problem by replacing the differential operators with differentiation matrices, thereby directly solving the differential equation by numerical interpolation by spectral methods. Advances in computational linear algebra allow this with ease  \cite{weideman2000matlab}, and have advantages over other matrix methods susceptible to the so called large f-d problem (e.g. the transfer matric method (TMM) and global matrix formalism \cite{thomson1950transmission,haskell1953dispersion,dunkin1965computation,knopoff1964matrix,Chaplain2021}). 

Assuming time-harmonicity, the governing equations \eqref{eq:waveeqns} can be written as the eigen-problems
\begin{equation}
    \begin{split}
        &\mathcal{L}\Phi = -\frac{\omega^2}{c_p^2}\Phi, \\ \\
        &\left[\left(\mathcal{L} - \frac{1}{r^2}\right)\Psi_{r} - \frac{2\mathrm{i}m}{r^2}\Psi_{\theta}\right]\boldsymbol{\hat{e}}_{r} = -\frac{\omega^2}{c_s^2}\Psi_{r}\boldsymbol{\hat{e}}_{r}, \\ \\
        &\left[\frac{2\mathrm{i}m}{r^2}\Psi_{r} + \left(\mathcal{L} - \frac{1}{r^2}\right)\Psi_{\theta} \right]\boldsymbol{\hat{e}}_{\theta} = -\frac{\omega^2}{c_s^2}\Psi_{\theta}\boldsymbol{\hat{e}}_{\theta}, \\ \\
        &\mathcal{L}\Psi_{z} = -\frac{\omega^2}{c_s^2}\Psi_{z},
    \end{split}
    \label{eq:operatoreqn}
\end{equation}
with
\begin{equation}
    \mathcal{L} = \left[\frac{\mathrm{d}^2}{\mathrm{d}r^2} + \frac{1}{r}\frac{\mathrm{d}}{\mathrm{d}r} - \left(\frac{m^2}{r^2} + k_{z}^2 \right) \right].
\end{equation}
Spectral collocation rests on representing this differential operator as a differentiation matrix. As outlined in \cite{adamou2004spectral}, these are introduced as matrices, $D^{(p)}$ which approximate the $p^{th}$ derivative of a column vector $\boldsymbol{f}$, whose entries are the values of the function $f(x)$ at the $N$ interpolation points $x_i$, $i = 1,\hdots, N$. Therefore the $p^{(th)}$ derivative is given by the matrix multiplication of $\boldsymbol{f}$ with the $N\times N$ matrix $D^{(p)}$ such that 
\begin{equation}
    \boldsymbol{f^{(p)}} \approx D^{(p)}\boldsymbol{f}.
\end{equation}
Given the bounded nature of the geometry in the radial direction, it is appropriate to use Chebyshev differentiation matrices \cite{trefethen2000spectral,boyd2001chebyshev}. These are evaluated using the MATLAB differentiation matrix suite \cite{weideman2000matlab} with a suitable coordinate transform to polar coordinates \cite{adamou2004spectral}. Therefore $\mathcal{L}$ can be represented, with sufficient ease, by the $N\times N$ matrix 
\begin{equation}
    L = D^{(2)} + \text{diag}\left(\frac{1}{r}\right)D^{(1)} - \text{diag}\left(\frac{m^2}{r^2} + k_{z}^2 \right).
\end{equation}
Following this \eqref{eq:operatoreqn} can be written as a matrix eigenvalue equation:
\begin{equation}
    {P}\bu = -\omega^2Q\bu,
\end{equation}
with $\bu = \left(\Phi(r_i),\Psi_{r}(r_{i}),\Psi_{\theta}(r_{i}),\Psi_{z}(r_{i})\right)^{T}$ where $i = 1,\hdots,N$, and 
\begin{equation}
    Q = \begin{pmatrix}
    M(c_p) &0 &0 &0 \\
    0 & M(c_s) &0 &0\\
    0 & 0& M(c_s) & 0\\
    0 & 0 &0 & M(c_s)
    \end{pmatrix},
\end{equation}
where $M(x)$ is the $N\times N$ matrix
\begin{equation}
    M(x) = \text{diag}\left(\frac{1}{x^2}\right).
\end{equation}
The matrix $P$ encodes the differential operators such that 
\begin{equation}
    P = \begin{pmatrix}    
    L & 0& 0& 0\\
    0 &L_1 & -L_2 & 0 \\
    0 & L_2 & L_1 & 0 \\
    0 & 0 & 0 & L
    \end{pmatrix}.
\end{equation}
where 
\begin{equation}
\begin{split}
    L_1 &= \left[L-\text{diag}\left(\frac{1}{r^2}\right)\right], \\ \\
    L_2 &= 2\mathrm{i}m\left[ \text{diag}\left(\frac{1}{r^2}\right)\right]
\end{split}
\end{equation}
The eigenvalue problem is completed by incorporating the traction free boundary conditions and the gauge condition in these matrices. This involves constructing the boundary condition matrix $S$ below, and replacing rows of $P$ and $Q$ to reflect the boundary conditions.

In terms of the elastic potential fields, the traction free conditions read
\begin{equation}
    \begin{split}
        {\sigma_{rr}} &= \lambda\nabla^2\Phi + 2\mu\frac{\mathrm{\partial} u_{r}}{\mathrm{\partial} r} \\ &= \left[\kappa^2\frac{\mathrm{d}^2}{\mathrm{d} r^2} +(\kappa^2-2)\left(\frac{1}{r}\frac{\mathrm{d}}{\mathrm{d} r} - \frac{m^2}{r^2} -k_{z}^2 \right) \right]\phi \\
        &-\left(2\mathrm{i}k_{z}\frac{\mathrm{d}}{\mathrm{d}r}\right)\psi_{\theta} + 2\mathrm{i}m\left(\frac{1}{r}\frac{\mathrm{d}}{\mathrm{d}r} -\frac{1}{r^2}\right)\psi_{z},
    \end{split}
    \label{eq:BC1}
\end{equation}
\begin{equation}
    \begin{split}
        {\sigma_{r\theta}} &= 2\mu\epsilon_{r\theta} = \mu\left[r\frac{\mathrm{\partial}}{\mathrm{\partial} r}\left(\frac{u_{\theta}}{r}\right) + \frac{1}{r}\frac{\mathrm{\partial} u_{r}}{\mathrm{\partial}\theta} \right] \\
        &= 2\mathrm{i}m\left(\frac{1}{r}\frac{\mathrm{d}}{\mathrm{d}r} -\frac{1}{r^2}\right)\phi + \mathrm{i}k_{z}\left(\frac{\mathrm{d}}{\mathrm{d}r} -\frac{1}{r}\right)\psi_{r} \\
        &+ \frac{k_{z}m}{r}\psi_{\theta} + \left(-\frac{\mathrm{d}^2}{\mathrm{d}r^2} + \frac{1}{r}\frac{\mathrm{d}}{\mathrm{d}r} - \frac{m^2}{r^2} \right)\psi_{z}
    \end{split}
    \label{eq:BC2}
\end{equation}
\begin{equation}
    \begin{split}
        {\sigma_{rz}} &= 2\mu\epsilon_{rz} = \mu\left(\frac{\mathrm{\partial} u_r}{\mathrm{\partial} z} + \frac{\mathrm{\partial} u_z}{\mathrm{\partial} r} \right) \\
        &= \left(2\mathrm{i}k_{z} \frac{\mathrm{d}}{\mathrm{d}r}\right)\phi + \mathrm{i}m\left(\frac{1}{r^2} - \frac{1}{r}\frac{\mathrm{d}}{\mathrm{d}r} \right)\psi_{r} \\
        &+ \left(\frac{\mathrm{d}^2}{\mathrm{d}r^2} + \frac{1}{r}\frac{\mathrm{d}}{\mathrm{d}r}  - \frac{1}{r^2} +k_{z}^2\right)\psi_{\theta} - \frac{m k_{z}}{r}\psi_{z},
    \end{split}
    \label{eq:BC3}
\end{equation}
where $\kappa = c_p/c_s$. The gauge condition then has the form
\begin{equation}
    \begin{split}
        \nabla\cdotp\boldsymbol{\Psi} = \left(\frac{1}{r} + \frac{\mathrm{d}}{\mathrm{d}r}\right)\psi_{r} + \frac{\mathrm{i}m}{r}\psi_{\theta} + \mathrm{i}k_{z}\psi_{z} = 0.
    \end{split}
    \label{eq:BC4}
\end{equation}
The vector we define as $\boldsymbol{s} = \left(\sigma_{rr}(r_{i})/\mu,\sigma_{r\theta}(r_{i})/\mu,\sigma_{rz}(r_{i})/\mu,\nabla\cdotp\Psi(r_i)\right)^T$ is then related to the displacement vector $\boldsymbol{u}$ through the $4N\times4N$ boundary condition matrix $S$ defined by
\begin{equation}
    S = \begin{pmatrix}
    S_{11}^{rr} & S_{12}^{rr} & S_{13}^{rr} & S_{14}^{rr} \\ \\
    S_{21}^{r\theta} & S_{22}^{r\theta} & S_{23}^{r\theta} & S_{24}^{r\theta} \\ \\
    S_{31}^{rz} & S_{32}^{rz} & S_{33}^{rz} & S_{34}^{rz} \\ \\
    S_{41}^{\nabla} & S_{42}^{\nabla} & S_{43}^{\nabla} & S_{44}^{\nabla}
    \end{pmatrix},
\end{equation}
such that $\boldsymbol{s} = S\bu$ and the components of $S$ are the $N\times N$ matrices determined by incorporating the differentiation matrices $D^{(p)}$ into equations \eqref{eq:BC1}-\eqref{eq:BC4}:
\begin{equation}
    \begin{split}
        S_{11}^{rr} &= \kappa^2D^{(2)} + (\kappa^2-2)\left[\text{diag}\left(\frac{1}{r}\right)D^{(1)} -m^2\text{diag}\left(\frac{1}{r^2} + k_z^2\right) \right], \\
        S_{12}^{rr} &= Z, \\
        S_{13}^{rr} &= -2\mathrm{i}\text{   diag}\left(k_z\right)D^{(1)}, \\
        S_{14}^{rr} &= 2\mathrm{i}m\left[\text{diag}\left(\frac{1}{r}\right)D^{(1)} - \text{diag}\left(\frac{1}{r^2}\right) \right], \\ 
        S_{21}^{r\theta} &= 2\mathrm{i}m\left[\text{diag}\left(\frac{1}{r}\right)D^{(1)} - \text{diag}\left(\frac{1}{r^2}\right) \right],\\
        S_{22}^{r\theta} &= \mathrm{i}\text{   diag}\left(k_z\right)\left[D^{(1)} - \text{diag}\left(\frac{1}{r}\right)\right],\\
        S_{23}^{r\theta} &= m\text{   diag}{\left(\frac{k_z}{r}\right)}, \\
        S_{24}^{r\theta} &= -D^{(2)} + \text{diag}{\left(\frac{1}{r}\right)D^{(1)}} - m^2\text{diag}\left(\frac{1}{r^2} \right),\\
        S_{31}^{rz} &= 2\mathrm{i}\text{   diag}\left(k_z\right)D^{(1)},\\
        S_{32}^{rz} &= \mathrm{i}m\left[\text{diag}\left(\frac{1}{r^2} \right) - \text{diag}\left(\frac{1}{r}\right)D^{(1)} \right],\\
        S_{33}^{rz} &= D^{(2)} + \text{diag}\left(\frac{1}{r}\right)D^{(1)} - \text{diag}\left(\frac{1}{r^2} - k_z^2 \right),\\
        S_{34}^{rz} &= -m\text{   diag}{\left(\frac{k_z}{r} \right)},\\
        S_{41}^{\nabla} &= Z,\\
        S_{42}^{\nabla} &= \text{diag}\left(\frac{1}{r}\right) + D^{(1)} ,\\
        S_{43}^{\nabla} &= \mathrm{i}m\text{   diag}\left(\frac{1}{r}\right),\\
        S_{44}^{\nabla} &= \mathrm{i}\text{   diag}\left(k_{z} \right),\\
    \end{split}
\end{equation}
where $Z$ is an $N\times N$ matrix of zeros. 

The boundary conditions are the built in to the eigenvalue problem by augmenting the rows of $P$ with the rows of $S$ that pertain to the coordinates for $r_{a}$ and $r_{b}$, namely $r_i$ and $r_N$, and enforcing their values by modifying $Q$. The spectral method is then fully described by the generalised eigenvalue problem
\begin{equation}
    \hat{P}\bu = -\omega^2\hat{Q}\bu, 
\end{equation}
where $\hat{P}$ is the matrix $P$, but with the $1^{st}$, $N^{th}$, $N+1^{th}$,$2N^{th}$,$2N+1^{th}$,$3N^{th}$,$3N+1^{th}$ and $4N^{th}$ rows replaced by the corresponding rows of $S$, and 
\begin{equation}
    \hat{Q} = \begin{pmatrix}
    \hat{M} &0 &0 &0 \\
    0 & \hat{M} & 0&0\\
    0&0&\hat{M}&0\\
    0&0&0&\hat{M}
    \end{pmatrix}M,
\end{equation}
such that $\hat{M}$ is the $N\times N$ matrix
\begin{equation}
    \hat{M} = \begin{pmatrix}
    0 &  & & &\\ 
     & 1 &  & &\\
    & & \ddots & &\\
    & & & 1 &\\
    & & & & 0
    \end{pmatrix}.
\end{equation}
This equation is solved numerically giving the dispersion relation $\omega(k_z)$ for a given circumferential order $m$. A pictorial representation of $\hat{P}$ is shown below.
\begin{figure}[h]
\centering
\includegraphics[width = 0.45\textwidth]{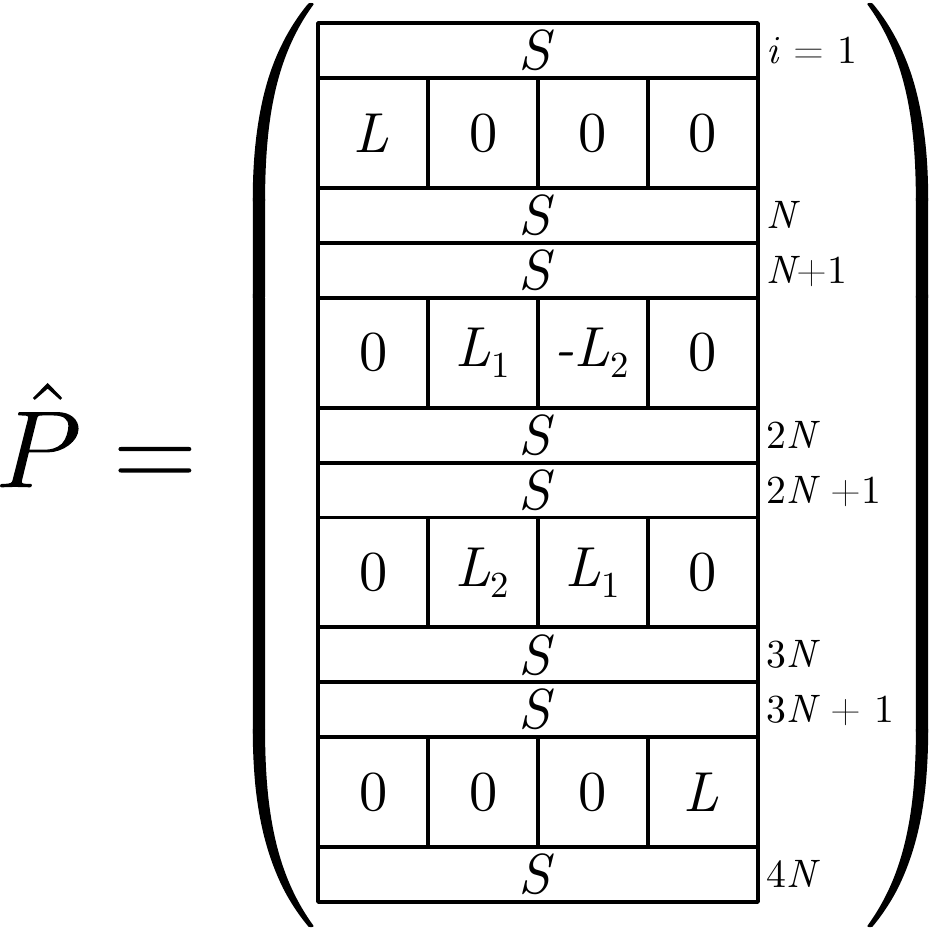}.
\end{figure}

The mode profiles can then be evaluated through each displacement component:
\begin{equation}
    \begin{split}
        u_{r} &= \frac{\mathrm{\partial}\Phi}{\mathrm{\partial} r}  + \frac{\mathrm{i}m}{r}\Psi_{z} - \mathrm{i}k_{z}\Psi_{\theta} \\ 
        u_{\theta} &= \frac{\mathrm{i}m}{r}\Phi + \mathrm{i}k_{z}\Psi_{r} - \frac{\mathrm{\partial}\Psi_{z}}{\mathrm{\partial} r} \\
        u_{z} &= \mathrm{i}k_{z}\Phi + \frac{\Psi_{\theta}}{r} + \frac{\mathrm{\partial}\Psi_{\theta}}{\mathrm{\partial} r} - \frac{\mathrm{i}m}{r}\Psi_{r}. \\
    \end{split}
\end{equation}

We note that using the displacement potentials is most appropriate for this simple example, but it is more appropriate to reformulate the SCM using the displacement fields and elasticity tensor coefficients to deal with anisotropy and more complex geometries, as the gauge and boundary conditions are difficult to incorporate otherwise.

\section{Supplementary Results}
\label{sec:appendB}
Here we shown an additional numerical FFT of a frequency domain FEM simulation for an extended pipe of length $10~\si{\meter}$. The eSPP is the same as in Figure~\ref{fig:Setup}. A harmonic forcing is applied on the end cap at each frequency and the FFT extracted along a line on the outer radius of the extended pipe after the eSPP. The dispersion curves for an infinitely long pipe are overlaid, just as in Figure~\ref{fig:numFFT}.
\begin{figure}[ht!]
    \centering
    \includegraphics[width = 0.475\textwidth]{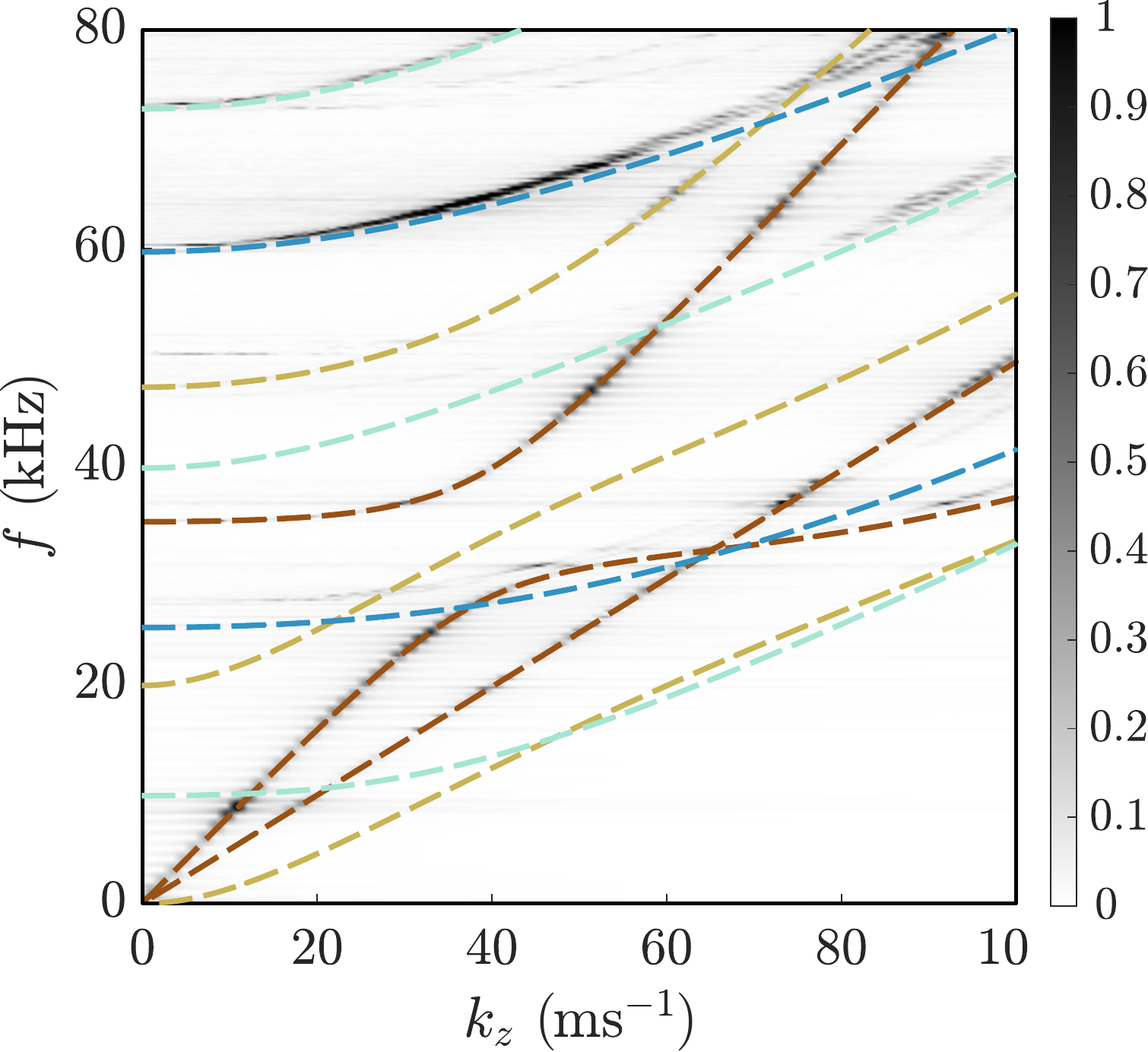}
    \caption{\textbf{FEM frequency domain FFT:} Simulation of extended pipe to improve resolution in $k_z$. Dispersion curves from SCM are also shown, highlighting the accuracy of the method for predicting the modes supported by the finite pipe.}
    \label{fig:SuppFFT}
\end{figure}


\end{document}